\documentclass[aps,floatfix]{revtex4}
 \usepackage[pdftex]{color,graphicx}

 \newcommand{\be}{\begin{equation}}
 \newcommand{\ee}{\end{equation}}
 \newcommand{\bea}{\begin{eqnarray}}
 \newcommand{\eea}{\end{eqnarray}}
 \newcommand{\mPl}{m_{\rm Pl}}
 \newcommand{\fNL}{f_{\rm NL}}
 \newcommand{\gNL}{g_{\rm NL}}
 \renewcommand{\P}{{\cal P}}
 \newcommand{\R}{{R_\chi}}
 \newcommand{\rT}{{r_T}}
 \newcommand{\osc}{{\rm osc}}
 \newcommand{\pGupta}{p}
 \newcommand{\pLW}{p_{\rm LW}}
 \newcommand{\pFW}{p_{\rm FW}}
  \renewcommand{\k}{{\vec k}}

\begin{document}

 {\hfill YITP-10-27}

\title{Non-Gaussianity and Gravitational Waves from Quadratic and Self-interacting Curvaton}

\author{Jos\'e Fonseca and David Wands}

\affiliation{Institute of Cosmology \& Gravitation, Dennis Sciama Building, Burnaby Road
University of Portsmouth, Portsmouth, PO1~3FX, United Kingdom\\
and\\
Yukawa Institute for Theoretical Physics, Kyoto University, Kyoto, 606-8502, Japan}

\date{January 18th, 2011}

\begin{abstract}
In this paper we consider how non-Gaussianity of the primordial density perturbation and the amplitude of gravitational waves from inflation can be used to determine parameters of the curvaton scenario for the origin of structure. We show that in the simplest quadratic model, where the curvaton evolves as a free scalar field, measurement of the bispectrum relative to the power spectrum, $\fNL$, and the tensor-to-scalar ratio can determine both the expectation value of the curvaton field during inflation and its dimensionless decay rate relative to the curvaton mass.
%
We show how these predictions are altered by the introduction of self-interactions, in models where higher-order corrections are determined by a characteristic mass scale and discuss how additional information about primordial non-Gaussianity and scale dependence may constrain curvaton interactions.
\end{abstract}
\maketitle


\section{Introduction}


Inflation solves the horizon problem, the flatness problem and the monopole problem. Furthermore, it gives a simple way to source primordial perturbations from quantum vacuum fluctuations. Any light scalar field during a period of inflation with an almost constant Hubble expansion acquires an almost scale-invariant power spectrum of fluctuations that could be the origin of primordial density perturbations \cite{Bassett:2005xm,Lyth:2009zz}.

The curvaton is one such field which is only weakly coupled and hence decays on a time-scale much longer than the duration of inflation \cite{Linde:1996gt,Enqvist:2001zp,Lyth:2001nq,Moroi:2001ct,Wands:2005aa}. Its lightness enables the field to acquire super-Hubble perturbations from vacuum fluctuations during inflation. When it decays into radiation some time after inflation has ended, its decay can source the perturbations in the radiation density of the universe, and all other species in thermal equilibrium \cite{Lyth:2002my,Weinberg:2003sw}.

One of the distinctive predictions of the curvaton scenario for the origin of structure is the possibility of non-Gaussianity in the distribution of the primordial density perturbations \cite{Bartolo:2004ty,Boubekeur:2005fj,Lyth:2005fi}. Treating the curvaton as a pressureless fluid one can estimate the resulting non-Gaussianity either analytically by treating the decay of the curvaton as instantaneous \cite{Lyth:2002my,Bartolo:2004ty,Lyth:2005fi}, or numerically \cite{Malik:2006pm,Sasaki:2006kq}, showing that the non-Gaussianity parameter $\fNL$ becomes large when the curvaton density at the decay time becomes small.

The non-linear evolution of the field before it decays can also contribute to the non-Gaussianity of the final density perturbation.
The authors of \cite{Enqvist:2005pg,Enqvist:2008gk,Huang:2008zj,Enqvist:2009zf,Enqvist:2009eq,Enqvist:2009ww} look at the effect of polynomial corrections to the quadratic curvaton potential. In some cases the curvaton density can be significantly subdominant at decay and still yield small $\fNL$ \cite{Enqvist:2009zf}. For small values of $\fNL$, the non-Gaussianity can instead be probed by the trispectrum parameter, $\gNL$.

Primordial gravitational waves on super-Hubble scales are also present since they are an inevitable byproduct at some level of an inflationary expansion. Non-Gaussianity alone could distinguish between the curvaton scenario and the conventional inflaton scenarios for the origin of structure since a single inflaton field is not capable of sourcing significant non-Gaussianity \cite{Maldacena:2002vr}. But non-Gaussianity and gravitational waves together can give tight constraints on curvaton model parameters. Nakayama {\em et al} \cite{Nakayama:2009ce} recently studied the effects of the entropy released by the decay of a curvaton field with a quadratic potential on the spectrum of gravitational waves that are already sub-horizon scale at the decay and consider the possibilities of future direct detection experiments, such as BBO or DECIGO, to constrain the parameter space. We will restrict our attention to gravitational waves on super-Hubble scales when the curvaton decays which are not affected by the decay, and consider self-interactions of the curvaton field in addition to the quadratic potential \cite{Enqvist:2010dt,Huang:2008zj}. This includes scales which contribute to the observed CMB anisotropies, where the power in gravitational waves is typically given by the tensor-to-scalar ratio for the primordial metric perturbations, $\rT$.

In this paper we will investigate how non-Gaussianity and gravitational waves provide constraints on curvaton model parameters. For any value of the curvaton model parameters we can obtain the observed amplitude of primordial density perturbations on large scales by adjusting the Hubble scale of inflation, which we assume to be an independent parameter in the curvaton model. However observational constraints on the tensor-to-scalar ratio places an upper bound on the inflationary Hubble scale, while non-Gaussianity constrains the remaining model parameters.


We numerically solve the evolution of the curvaton field in a homogeneous radiation-dominated era after inflation allowing for non-linear evolution of the curvaton field due to both explicit self-interaction terms in the potential and the self-gravity of the curvaton. In particular we consider quadratic and non-quadratic potentials which reduce to a quadratic potential about the minimum with self-interaction terms governed by a characteristic mass scale, corresponding to cosine or hyperbolic-cosine potentials. Cosine potentials arise for PNGB axion fields and are often considered as candidate curvaton fields \cite{Chun:2004gx,Dimopoulos:2005bx,Kawasaki:2008mc,Chingangbam:2009xi,Huang:2010cy}. The hyperbolic cosine is representative of a potential where self-interaction terms become large beyond a characteristic scale. In each case we show how the non-linearity parameter $\fNL$ and tensor-to-scalar ratio, $\rT$, can be used to determine model parameters.

In Section~II we review the perturbations generated during inflation and how these are transfered to the primordial density perturbation in the curvaton scenario. In Section~III we present the numerical results of our study for three different curvaton potentials. We conclude in Section~IV.

\section{Inflationary perturbations in the curvaton scenario}

In the curvaton scenario, initial quantum fluctuations in the curvaton field, $\chi$, during a period of inflation at very early times give rise to the primordial density perturbation in the subsequent radiation-dominated universe some time after inflation and after the curvaton field has decayed into radiation, e.g., the density perturbation in the epoch of primordial nucleosynthesis. This primordial density perturbation is conveniently characterised by the gauge-invariant variable, $\zeta$, corresponding to the curvature perturbation on uniform-density hypersurfaces \cite{Malik:2008im}.

Throughout this paper we will use the $\delta N$ formalism \cite{Starobinsky:1986fxa,Sasaki:1995aw,Wands:2000dp,Lyth:2005fi} to compute the primordial density perturbation in terms of the perturbation in the local integrated expansion, $N$, from an initial spatially-flat hypersurface during inflation, to a uniform-density hypersurface in the radiation-dominated era
\be \label{zetadeltachi}
\zeta=\delta N=N'\delta\chi_*+\frac 12 N''\delta\chi_*^2+
 + \ldots
\ee
where $\delta\chi_*=\chi_*-\langle\chi_*\rangle$ and primes denote derivatives with respect to $\chi_*$, the local value of the curvaton during inflation.

Quantum fluctuations of a weakly-coupled field on super-Hubble scales ($k/a\ll H$) during slow-roll inflation is well described by a Gaussian random field with two-point function
\be
\langle \chi_{\k_1} \chi_{\k_2} \rangle = (2\pi)^3 P_{\chi}(k_1)\delta^3(\k_1+\k_2) \,.
\ee
We define the dimensionless power spectrum $\P_{\chi}(k)$ as
\be
\P_{\chi}(k) \equiv \frac{k^3}{2\pi^2} P_{\chi}(k)
\ee
The power spectrum of curvature perturbations is thus given, at leading order, by
\be
\P_{\zeta}(k)=N'^2\P_{\delta\chi}(k) \,.
\ee
and we define the spectral index as
\be
n_\zeta-1 \equiv \frac{d\ln \P_\zeta}{d\ln k} \,,
\ee
and the running of the spectral index as
\be
\label{alphazeta}
\alpha_\zeta \equiv \frac{d\ln |n_\zeta-1|}{d\ln k} \,.
\ee

The connected higher-order correlation functions are suppressed for a weakly-coupled scalar field during slow-roll inflation, but non-linearities in the dependence of $N$ and hence $\zeta$ on the initial curvaton value in Eq.~(\ref{zetadeltachi}) can lead to significant non-Gaussianity of the higher-order correlation functions, in particular the bispectrum
\bea
\langle \zeta_{\k_1} \zeta_{\k_2} \zeta_{\k_3}\rangle&=& (2\pi)^3 B_{\zeta}(k_1,k_2,k_3)\delta^3(\k_1+\k_2+\k_3)
 \,.
\eea
The bispectrum
is commonly expressed in terms of the dimensionless non-linearity parameter, $\fNL$,
such that
\bea
B_{\zeta}(k_1,k_2,k_3) =\frac 65 \fNL \left[ P_{\zeta}(k_1)P_{\zeta}(k_2)+P_{\zeta}(k_1)P_{\zeta}(k_3)+P_{\zeta}(k_2)P_{\zeta}(k_3) \right]
\eea

If the initial field perturbations, $\delta\chi_*$, correspond to a Gaussian random field then it follows from Eq.~(\ref{zetadeltachi}) that $\fNL$
is independent of the wavenumbers, $k_i$, and is given by
\bea
\fNL&=&\frac 56 \frac{N''}{N'^2} \,. \label{deffnl}
\eea
In practice non-linear evolution of the field can lead to non-Gaussianity of the field perturbations on large scales and a weak scale dependence of $\fNL$
\cite{Byrnes:2009pe,Byrnes:2010ft,Byrnes:2010xd}.


Current bounds from the CMB on local-type non-Gaussianity require $-10<\fNL<74$ \cite{Komatsu:2010fb}. Large-scale structure surveys lead to similar bounds \cite{Slosar:2008hx}.

\subsection{Isocurvature field perturbations during inflation}

Perturbations of an isocurvature field, whose fluctuations have negligible effect on the total energy density, can be evolved in an unperturbed FRW background and obey the wave equation
\begin{equation}
 \label{chieom}
 \ddot{\delta\chi}+3H\dot{\delta\chi} +\bigg(\frac{k^2}{a^2}+m_\chi^2 \bigg)\delta\chi=0
 \,,
\end{equation}
where the effective mass-squared is given by $m_\chi^2=\partial^2 V/\partial\chi^2$.
During any period of accelerated expansion quantum vacuum fluctuations on small sub-Hubble scales (comoving wavenumber $k>aH$) are swept up to super-Hubble scales ($k<aH$). For a light scalar field, $\chi$, with effective mass much less than the Hubble rate during inflation ($m_{\chi*}^2\ll H_*^2$) the power spectrum of fluctuations at Hubble exit is given by
 \be
 \label{H2pi}
 {\cal P}_{\chi_*} \simeq \left( \frac{H_*}{2\pi} \right)^2 \quad {\rm for}\ k=a_*H_* \,.
 \ee
On super-Hubble scales the spatial gradients can be neglected and the overdamped evolution (\ref{chieom}) for a light field is given by
 \be
 H^{-1} \dot{\delta\chi} \simeq - \eta_\chi \delta\chi \,.
 \ee
where we define the dimensionless mass parameter
\be
 \eta_\chi = \frac{m_{\chi}^2}{3H^2} \,.
 \ee
Combined with the time-dependence of the Hubble rate in Eq.~(\ref{H2pi}), given by the slow-roll parameter $\epsilon \equiv -\dot{H}/H^2$, this
leads to a scale-dependence at any given time of the field fluctuations on super-Hubble scales \cite{Lyth:2001nq,Wands:2002bn}
 \be
 \label{nchi}
\Delta n_\chi \equiv \frac{d}{d\ln k} {\cal P}_\chi \simeq -2\epsilon + 2 \eta_\chi \,.
 \ee
which is small during slow-roll inflation, $\epsilon\ll1$, for light fields with $|\eta_\chi|\ll1$.

Self-interaction terms in the curvaton potential during inflation only modify the predictions for the power spectrum and spectral tilt beyond these leading order results in the slow-roll approximation. However they do lead to time-dependence of the effective mass of the $\chi$ field, so that the effective mass appearing in the expression for the spectral tilt may differ from that when the curvaton oscillates about the minimum of its potential some time after inflation. In particular the effective mass-squared during inflation could be negative, leading to a negative tilt, $\Delta n_\chi<0$, even if $\epsilon$ is very small.

The time-dependence of both $\epsilon$ and $\eta_\chi$
\begin{eqnarray}
 \label{dotetachi}
 H^{-1} \dot{\eta_{\chi}} &\simeq& 2 \epsilon{\eta_{\chi}} - {\xi^2_{\chi\phi}} \\
 \label{dotepsilon}
 H^{-1} \dot{\epsilon} &\simeq& -2\epsilon ({\eta_{\phi}}-2\epsilon) \label{derslow}
\end{eqnarray}
during slow-roll inflation driven by an inflaton field with dimensionless mass $\eta_\phi=V_{\phi\phi}/3H^2$ and $\xi_{\chi\phi}^2=(\partial^4V/\partial\chi^3\partial\phi)/9H^4$, gives rise to a running of the spectral index in Eq.~(\ref{nchi}) \cite{Wands:2003pw}
 \be
 \alpha_\chi \equiv \frac{d\ln \Delta n_\chi}{d\ln k} \simeq 4\epsilon \left( -2\epsilon + \eta_\phi + \eta_\chi \right) - 2\xi_{\chi\phi}^2 \,,
 \ee
In the following we shall make the usual assumption that the curvaton has no explicit interaction with the inflaton, so that $\xi_{\chi\phi}=0$ and the running is second-order in slow-roll parameters and expected to be very small. Note, however, that in the curvaton scenario the tensor-to-scalar ratio and spectral tilt do not directly constrain the slow-roll parameters $\epsilon$ and $\eta_\phi$ as in single-inflaton-field inflation, so they could be relatively large.

\subsection{Transfer to curvaton density}

In the curvaton scenario, these super-Hubble fluctuations in a weakly-coupled field whose energy density is negligible during inflation generates the observed primordial curvature perturbation, $\zeta$, after inflation if the curvaton comes to contribute a non-negligible fraction of the total energy density after inflation.

As the curvaton density becomes non-negligible one must include the backreaction of the field fluctuations on the spacetime curvature. However on super-Hubble scales, $k\ll aH$, where spatial gradients and anisotropic shear become negligible we can model the non-linear evolution of the field in terms of locally FRW dynamics \cite{Salopek:1990jq}. In the following we will employ this ``separate universe'' picture \cite{Wands:2000dp} and we have
\bea
 \label{chiL}
 \ddot{\chi}_L + 3H_L\dot\chi_L + V_{\chi L} \simeq 0 \,, \nonumber\\
 H_L^2 \simeq \frac{8\pi G}{3} \left( V_L + \frac12 \dot\chi_L^2 \right) \,.
 \eea
where $\chi_L=\chi+\delta\chi$, $H_L$, $V_L$ and $V_{\chi L}$ denote the field, Hubble rate, potential and potential gradient smoothed on some intermediate scale $(aH)^{-1}\ll L< k^{-1}$, and dots denote derivatives with respect to the local proper time.

Once the Hubble rate drops below the effective mass scale, the long-wavelength modes of the field, $\chi_L$, oscillate about the minimum of the potential. Any scalar field with finite mass has a potential which can be approximated by a quadratic sufficiently close to its minimum, and the effective equation of state, averaged over several oscillation times, becomes that of a pressureless fluid
 \begin{equation}
 \rho_\chi = \langle \frac12 m_\chi^2 \chi_L^2 + \frac12 \dot\chi_L^2 \rangle \propto a^{-3} \,.
 \end{equation}
Thus the energy density of the curvaton grows relative to radiation, $\rho_\gamma\propto a^{-4}$. The curvaton must eventually decay if it is to transfer its inhomogeneous density into a perturbation of the radiation density. We assume a slow, perturbative decay of the curvaton at a fixed decay rate, $\Gamma\ll m$ (though we note that oscillating fields can also undergo a non-perturbative decay, or partial decay at earlier times \cite{Enqvist:2008be,Chambers:2009ki}).

In this work we will numerically solve for the evolution of the curvaton field until it begins oscillating and determine its subsequent energy density. In order to follow the subsequent evolution and eventual decay of the curvaton density on time scales, $\sim \Gamma^{-1}$, much longer than the oscillation time, $\sim m^{-1}$, we adopt the results of Ref.~\cite{Malik:2002jb}.

Once the curvaton field behaves as a pressureless fluid, one can show that phase-space trajectory is determined by the dimensionless parameter \cite{Malik:2002jb,Gupta:2003jc}
 \be
 \label{pdef}
 \pGupta \equiv \lim_{\Gamma/H\to0} \Omega_{\chi} \sqrt{\frac{H}{\Gamma}} \,.
 \ee
In practice one can only treat the curvaton field as a pressureless fluid once it has begun to oscillate about the minimum of its potential. Taking the density of the curvaton when it begins to oscillate, $\rho_{\chi,\osc}\simeq m^2\chi^2_\osc/2$ in Eq.~(\ref{pdef}), we can estimate $\pGupta$ as \cite{Lyth:2001nq}
\begin{equation} \label{posc}
 \pGupta \simeq \pLW \equiv \frac{\chi^2_\osc}{6\mPl^2} \sqrt{\frac{m}{\Gamma}} .
\end{equation}
where the subscript ``osc'' denotes the time for which $H_\osc= m_\chi$ and $\mPl\equiv (8\pi G)^{-1/2}\simeq 2.43\times 10^{18}$GeV is the reduced Planck mass.
Although the actual time when the curvaton begins oscillating is also not precisely defined this need not be a problem as $\Omega_{\chi} \sqrt{{H}/{\Gamma}}$ is a constant while the curvaton is sub-dominant at early times, since $\Omega_\chi\propto a \propto t^{1/2}$ and $H\propto t^{-1}$ for a pressureless fluid in a radiation dominated era, and we simply require $\chi^2_\osc/6\mPl^2\ll 1$.

However, Eq.(\ref{pdef}) only estimates $p$ in terms of the curvaton field value when the curvaton starts oscillating and we have assumed it has a quadratic potential at this time. More generally, to allow for self-interactions of the curvaton field that could lead to non-linear evolution after inflation and could still be significant when the curvaton begins to oscillate we define a transfer function for the field $\chi_\osc = g(\chi_*)$ \cite{Sasaki:2006kq} such that
\be
 \label{defg}
 p \equiv \frac{g^2(\chi_*)}{6\mPl^2} \sqrt{\frac{m}{\Gamma}}
 \,.
 \ee
in order to relate the density of curvaton at late times, as it oscillates about the minimum of its potential, to the value of the curvaton field during inflation, $\chi_*$.

\subsection{Transfer to primordial perturbation}

The amplitude of the resulting primordial curvature perturbation depends both on the perturbation in the curvaton density, $\delta\rho_\chi/\rho_\chi$, and the energy density in the curvaton field when it decays.
To first-order in the perturbations we write
\be
 \label{Rchi}
 \zeta = \R \left( \frac{\delta\rho_\chi}{3\rho_\chi}\right)_{\rm osc} = \R \frac{\delta p}{3p} \,.
 \ee
where $0<\R<1$ is a dimensionless efficiency parameter related to the fraction of the total energy density in the curvaton field when it decays into radiation.
Using the separate universe picture, we take derivatives of the same function $g(\chi_*)$ defined in terms of the homogeneous background fields in Eq.~(\ref{defg}) to determine the linear density perturbation and higher-order perturbations in terms of the field perturbations during inflation.
We thus have the transfer function for linear curvaton field perturbations during inflation into the primordial curvature perturbation
\be
\label{zeta}
 \zeta = \R \frac{1}{3} \frac{p'\delta\chi_*}{p} = \R \frac{2}{3} \frac{g'\chi_*}{g} \frac{\delta\chi_*}{\chi_*} \,.
 \ee
where primes denote derivatives with respect to $\chi_*$.

Modelling the transfer of energy from the curvaton field to the primordial radiation by a sudden decay at a fixed value of $H_{\rm decay}=\Gamma$ gives the transfer parameter \cite{Lyth:2001nq,Lyth:2002my}
 \be
 \R \approx \left[ \frac{3\rho_\chi}{4\rho_{\rm total}-\rho_\chi}\right]_{\rm decay} \,.
 \ee
However this expression is of limited use if we want to predict the primordial curvature perturbation in terms of the inflationary value of the curvaton field and its perturbations because this expression refers to the curvaton density at the decay time. The curvaton density changes with time and the decay time is not precisely defined since the decay happens over a finite period of time around $H\sim \Gamma$.

More generally, the transfer parameter, $\R$ in Eq.~(\ref{Rchi}), is a smooth function of the phase-space parameter $p$ defined in Eq.~(\ref{pdef}).
One can determine $\R$ as a function of $\pGupta$ numerically, which gives the analytic approximation \cite{Gupta:2003jc}
\begin{equation}
 \label{Rp}
 \R(\pGupta) \simeq 1 - \bigg( 1+\frac{0.924}{1.24}\pGupta \bigg)^{-1.24} .
\end{equation}


A distinctive feature of the curvaton scenario is the possibility that the primordial curvature perturbation may have a significantly non-Gaussian distribution even if the curvaton field itself is well described by a Gaussian distribution. This is due primarily to the fact that the energy density of a massive scalar field when it oscillates about the minimum of its potential is a quadratic function of the field. Simply assuming a linear transfer (\ref{Rchi}) from a quadratic curvaton density to radiation yields \cite{Lyth:2002my}
\be
 \zeta = \frac{\R}{3} \left( \frac{2\chi\delta\chi + \delta\chi^2}{\chi^2} \right) \,,
 \ee
and hence a primordial bispectrum of local form \cite{Komatsu:2001rj} characterised by the dimensionless parameter
\be \label{fnlrsmall}
 \fNL = \frac{5}{4\R} \,.
  \ee
This provides a good estimate of the non-Gaussianity for a quadratic curvaton with Gaussian distribution when $\fNL\gg1$.

Incorporating the full non-linear transfer for a quadratic curvaton density while assuming the curvaton field has a Gaussian distribution at a sudden decay, yields corrections of order unity \cite{Bartolo:2004ty,Lyth:2005fi,Sasaki:2006kq}
 \be
 \label{fNLRlinear}
 \fNL \simeq \frac{5}{4\R} -\frac53 -\frac{5\R}{6} \,.
 \ee
Numerical studies \cite{Malik:2006pm,Sasaki:2006kq} confirm that this sudden-decay formula for $\fNL(\R)$ represents an excellent approximation to the actual exponential decay, $n_\chi\propto e^{-\Gamma t}/a^3$, where we take $\R$ in Eq.~(\ref{fNLRlinear}) to be the linear transfer efficiency defined by Eq.~(\ref{Rchi}). In particular we find the robust result $\fNL\geq-5/4$ for any value of $\R$.

More generally, if we allow for possible non-linear evolution of the local curvaton field after Hubble-exit through the function $g(\chi_*)$ defined in Eq.~(\ref{defg}), and allow for possible variation of the transfer parameter $\R$ with the value of the curvaton field (but still take the curvaton fluctuations to be Gaussian at Hubble-exit) then we have \cite{Sasaki:2006kq}
 \be
 \label{fNLRused}
 \fNL = \frac{5}{4\R} \left[\left( 1 + \frac{gg''}{g^{\prime2}} \right)+\frac{\R' \left(g/g'\right)-2\R}{\R}\right]
 \, .
 \ee
This expression follows directly from Eq.~(\ref{deffnl}) when we take $N'=\frac 23 \R \frac{g'}{g}$.

If we adopt the sudden-decay approximation for $\R(p)$ then Eq.~(\ref{fNLRused}) reduces to~\cite{Lyth:2005fi}
 \be
 \label{fNLR}
 \fNL \simeq \frac{5}{4\R} \left( 1 + \frac{g''g}{g^{\prime2}} \right) -\frac53 -\frac{5\R}{6} \,.
 \ee

\subsection{Metric perturbations during inflation}

In most studies of the curvaton scenario it is assumed that the amplitude of scalar or metric perturbations generated during inflation are completely negligible. Indeed the original motivation for the study of the curvaton was to show that it was possible for fluctuations in a field other than the inflaton to completely dominate the primordial curvature perturbation. However gravitational waves describe the free oscillations of the metric tensor,  independent (at first order) of the matter perturbations, and some amplitude of fluctuations on super-Hubble scales is inevitably generated during an accelerated expansion. The resulting power spectrum of tensor metric perturbations is given by
\begin{equation}
 \label{powergw}
 \P_T = \frac{8}{\mPl^2} \bigg( \frac{H_*}{2 \pi} \bigg)^2 \,.
 \end{equation}
The power spectrum of primordial gravitational waves if they can be observed by future CMB experiments, such as CMBPol \cite{Baumann:2008aq}, would give a direct measurement of the energy scale of inflation and hence the Hubble rate, $H_*$.
In practice the amplitude of gravitational waves is usually expressed relative to the observed primordial curvature perturbation as the tensor-to-scalar ratio
 \begin{equation} \label{ttsratio}
 \rT
  \equiv \frac{\P_T}{\P_\zeta}
    \simeq 8.1 \times 10^7 \bigg( \frac{H_*}{\mPl}\bigg)^2
 = 0.14 \times \bigg( \frac{H_*}{10^{14} {\rm GeV}} \bigg)^2
 \,.
 \end{equation}
Current observational bounds from CMB anisotropies are partially degenerate with bounds on the spectral index and dependent on theoretical priors, but can be used give $\rT<0.24$ \cite{Komatsu:2010fb}. Bounds from the power spectrum of the B-mode polarisation of the CMB are less model dependent and require $\rT<0.72$ \cite{Chiang:2009xsa}.

The tensor perturbations are massless and the scale dependence of the spectrum after Hubble-exit (\ref{powergw}) is simply due to the time dependence of the Hubble rate:
 \be
 \label{nT}
 n_T = -2\epsilon \,.
 \ee
Thus the tilt of the gravitational wave spectrum on very large scales today gives a direct measurement of the equation of state during inflation, $w=-1+(2\epsilon/3)$.

If inflation is driven by a light inflaton field, $\varphi$, this inflaton field also inevitably acquires a spectrum of fluctuations during the accelerated expansion, $\P_{\varphi_*} =(H/2\pi)_*^2$. These adiabatic field perturbations~\cite{Gordon:2000hv} correspond to a curvature perturbation at Hubble-exit during inflation
 \be
  \label{zetastar}
 \P_{\zeta*} = \left( \frac{H}{\dot\varphi} \right)_*^2 \P_{\varphi_*} = \frac{1}{16\epsilon} \P_T \,.
 \ee
The scale-dependence of the tensor spectrum (\ref{nT}) together with the time-dependence of $\epsilon$ during inflation, given in Eq.~(\ref{dotepsilon}), leads to a scale dependence of the curvature perturbation from adiabatic perturbations
 \be
 n_{\zeta_*}-1 = - 6\epsilon + 2\eta_\varphi \,,
 \ee
where the dimensionless inflaton mass parameter is $\eta_\varphi= m_\varphi^2/3H^2$. Note that the primordial curvature perturbation due to canonical inflaton field perturbations is effectively Gaussian with $|\fNL|_*\ll1$ suppressed by slow-roll parameters \cite{Maldacena:2002vr}.

In the presence of a curvaton field, the adiabatic perturbations during inflation represent only a lower bound on the primordial curvature perturbation and one should add the uncorrelated contributions to the primordial curvature perturbation from both the curvaton field (\ref{zeta}) and the inflaton field (\ref{zetastar}):
 \be
 \P_\zeta = \left( \frac{2g'\R}{3g} \right)^2 \P_\chi + \frac{1}{16\epsilon} \P_T \,.
 \ee

For example, if the spectral tilt of the primordial curvature perturbation from a very light curvaton field (\ref{nchi}) is $n_\chi-1\approx -0.03$ and primarily due to the time-dependence of the Hubble rate during inflation, $n_\chi-1\approx n_T \simeq -2\epsilon$, then we have $16\epsilon\approx 16\times0.015=0.24$ and hence $\P_{\zeta*}\approx 4\P_T$. Hence $\P_{\zeta*}\ll\P_\zeta$ for $\rT\ll0.3$.

In the following we will assume $\epsilon$ is large enough that the inflaton contribution to the primordial curvature perturbation can be neglected even if the primordial tensor perturbations are potentially observable.

\section{Numerical results}


In our numerical analysis we have used the separate universe equations (\ref{chiL}) to evolve the local value of $\chi_L$ for long-wavelength perturbations of the curvaton field. This incorporates both the non-linear self-interactions included in the potential of the curvaton, $V(\chi_L)$, and non-linearity of the gravitational coupling through the dependence of the Hubble expansion rate on the curvaton field kinetic and potential energy density in the Friedmann equation.

We do not solve for the curvaton field evolution during inflation or during (p)reheating at the end of inflation since this would be model dependent. Instead we start the evolution with a radiation density such that the initial Hubble rate is much larger than the effective mass of the curvaton, consistent with our assumption that the initial value of the curvaton field is effectively the same as its value at the end of inflation, $\chi_*$.

We evolve the curvaton until it begins to oscillate in the minimum of its potential and can accurately be described as a pressureless fluid, in order to exploit earlier work which used a fluid model to study the linear \cite{Gupta:2003jc} and non-linear \cite{Sasaki:2006kq} transfer of the curvaton perturbation to radiation and hence the primordial curvature perturbation. Thus we evolve the curvaton field until $\rho_\chi\propto a^{-3}$. Note that this may be sometime after the time when $H=m_\chi$ since the curvaton potential may have significant non-quadratic corrections at this time.

We need to be able to determine the dimensionless parameter $p$ defined in Eq.~(\ref{pdef}) which determines the transfer parameter $\R(\pGupta)$. To do so we identify
\be
 \label{pvpFW}
 \pGupta = \sqrt{\frac{m}{\Gamma}} \pFW \,.
 \ee
where
\be
 \label{pFW}
 \pFW \equiv \Omega_\chi (1-\Omega_\chi)^{-3/4} \sqrt{\frac{H}{m}} \,,
 \ee
is constant for a pressureless fluid, $\chi$, plus radiation.
It is straightforward to check that Eq.~(\ref{pvpFW}) coincides with the definition of $p$ given in Eq.~(\ref{pdef}), which is evaluated in the early time limit, $\Omega_\chi\to0$. The advantage of our variable $\pFW$ is that it can evaluated at late times, so long as the curvaton decay is negligible, $\Gamma\ll H$, whereas at early times the curvaton field may never actually evolve like a pressureless fluid and we may not have a well-defined early time limit for $\Omega_\chi\sqrt{H/\Gamma}$.

In our numerical code following the curvaton field evolution we use Eq.~(\ref{chiL}) with the rescaled time variable $\tau=mt$, implicitly setting $\Gamma=0$, such that
\bea
\chi'' + 3h \chi' +\frac{V_\chi}{m^2} = 0 \,,\\
h^2 = \frac{8\pi}{3\mPl^2} \left( \frac{V}{m^2} + \frac12 \chi^{\prime2} \right) \,.
\eea
For a quadratic potential we have $V_\chi/m^2=\chi$ and $V/m^2=\chi^2/2$ and hence the evolution of $\chi(\tau)$ is independent of $m$.
We evolve the curvaton field from an initial value $\chi_i=\chi_*$ when $H_i^2=100V_{\chi\chi}$. This is consistent with the usual assumption that the curvaton is a late-decaying field with $\Gamma\ll m$. We are then able to determine $\pFW(\chi_*)$ which approaches a constant as the curvaton density approaches that of a pressureless fluid at late times. We then obtain the actual parameter $\pGupta$ in Eq.~(\ref{pvpFW}) for a finite decay rate, by multiplying by a finite value of $\sqrt{m/\Gamma}$. Thus the parameter $\pGupta$ is a function of $\chi_*$ and $m/\Gamma$, but not $m$ and $\Gamma$ separately.

We use the previously determined \cite{Gupta:2003jc} transfer function $\R(\pGupta)$ given by Eq.~(\ref{Rp}). {}From Eq.~(\ref{H2pi}) and~(\ref{zeta}) we then have
 \be
  \label{Pzeta}
 \P_\zeta = \left( \frac{p'\R(p)}{3p} \right)^2 \left( \frac{H_*}{2\pi} \right)^2 \,.
 \ee
Normalising the amplitude of the primordial power spectrum to match the observed value on CMB scales, $\P_\zeta \simeq 2.5 \times 10^{-9}$ \cite{Komatsu:2010fb}, then fixes the amplitude of vacuum fluctuations of the curvaton field during inflation and hence the scale of inflation
 \begin{equation} \label{hp}
 H_*= 9.4 \times 10^{-4} \left( \frac{p}{p'\R(p)\mPl} \right) \mPl \, .
 \end{equation}
 or, equivalently, the tensor-scalar ratio
 \begin{equation}
 \label{rTp}
\rT
 = 72 \left( \frac{p}{p'\R(p)\mPl} \right)^2 \, .
 \end{equation}

The non-linearity parameter, $\fNL$, is given by Eq.~(\ref{fNLRused}). Note that for $\rT$ we must determine not only $p$ but also its first derivative, $p'$, with respect to the initial field value, $\chi_*$. For the non-linearity parameter, $\fNL$, we also need the second derivative, $p''$, and to describe higher-order non-Gaussianity we would need higher derivatives. In terms of the parameter $p$, Eq.~(\ref{fNLRused}) becomes
 \be
 \fNL = \frac{5}{2\R} \left[\frac{pp''}{p'^2} + \frac{\R'}{\R} \frac{p}{p'}-1\right] \,.
 \ee

\subsection{Quadratic curvaton}

\begin{figure}
\centering
\includegraphics[width=0.5\textwidth]{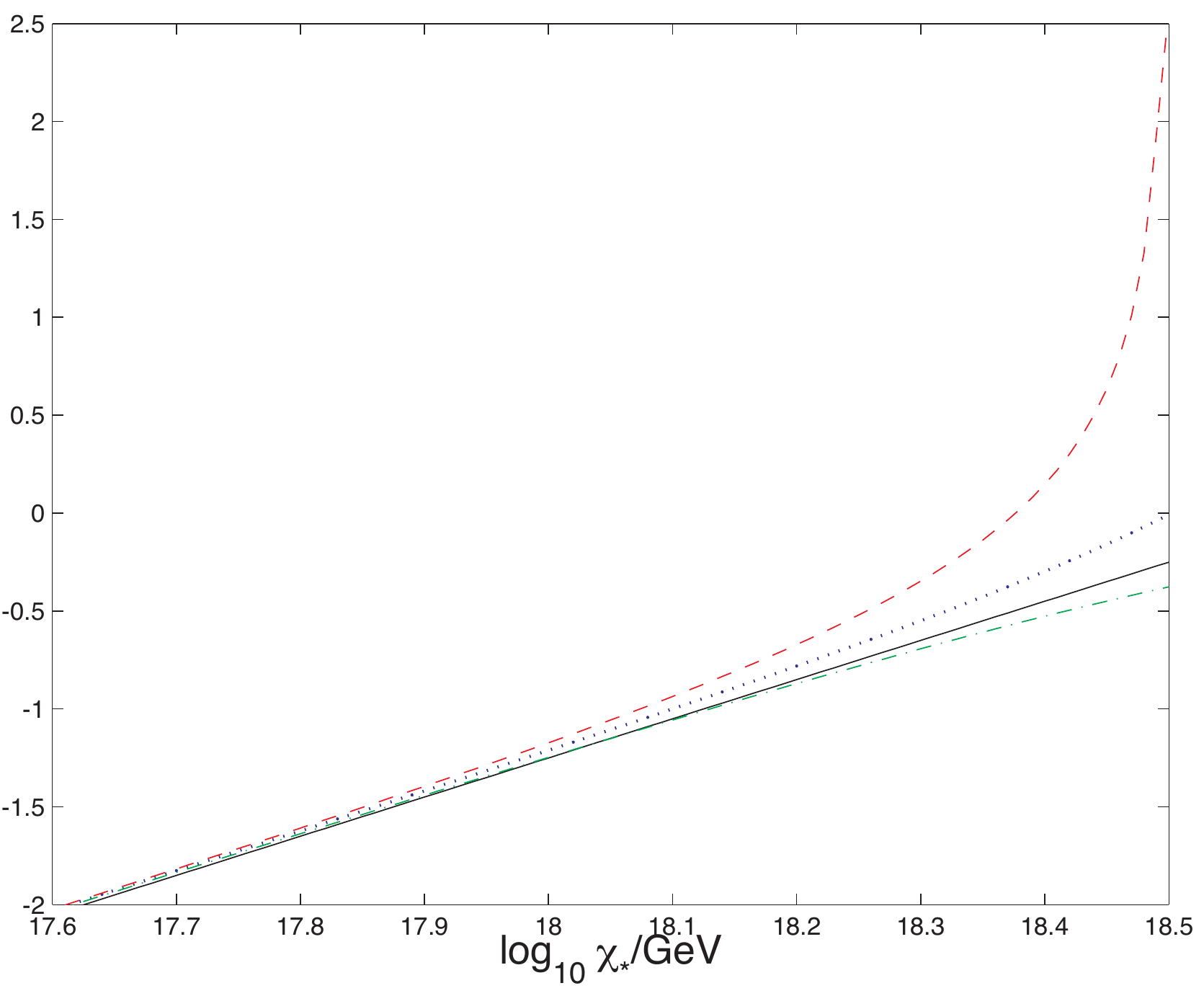}
\caption{Dimensionless curvaton parameter $p_{FW}$, defined in Eq.~(\ref{pFW}) as a function of initial curvaton field value, $\chi_*$, for three different potentials: quadratic potential (dotted blue line), cosine potential with $f=10^{18} \rm GeV$ (upper red dashed line) and hyperbolic cosine potential with $f=10^{18} \rm Gev$ (lower green dot-dashed line). For comparison, the solid black line shows $\chi_*^2/3\mPl^2$, which provides an excellent approximation for $\chi_*\ll \mPl$.}
%
\label{compq}
\end{figure}

\begin{figure}
\centering
\includegraphics[width=0.5\textwidth]{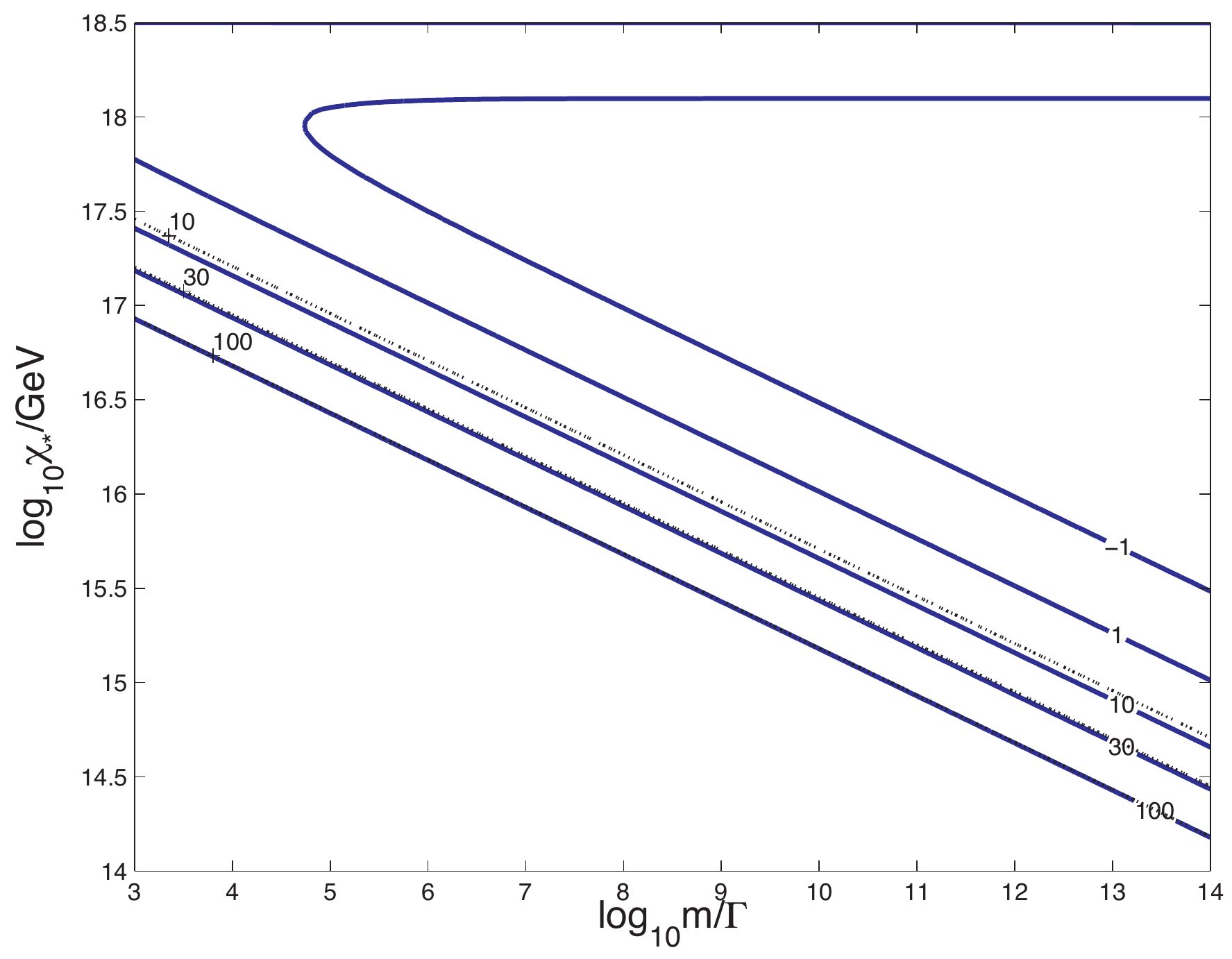}
\includegraphics[width=0.5\textwidth]{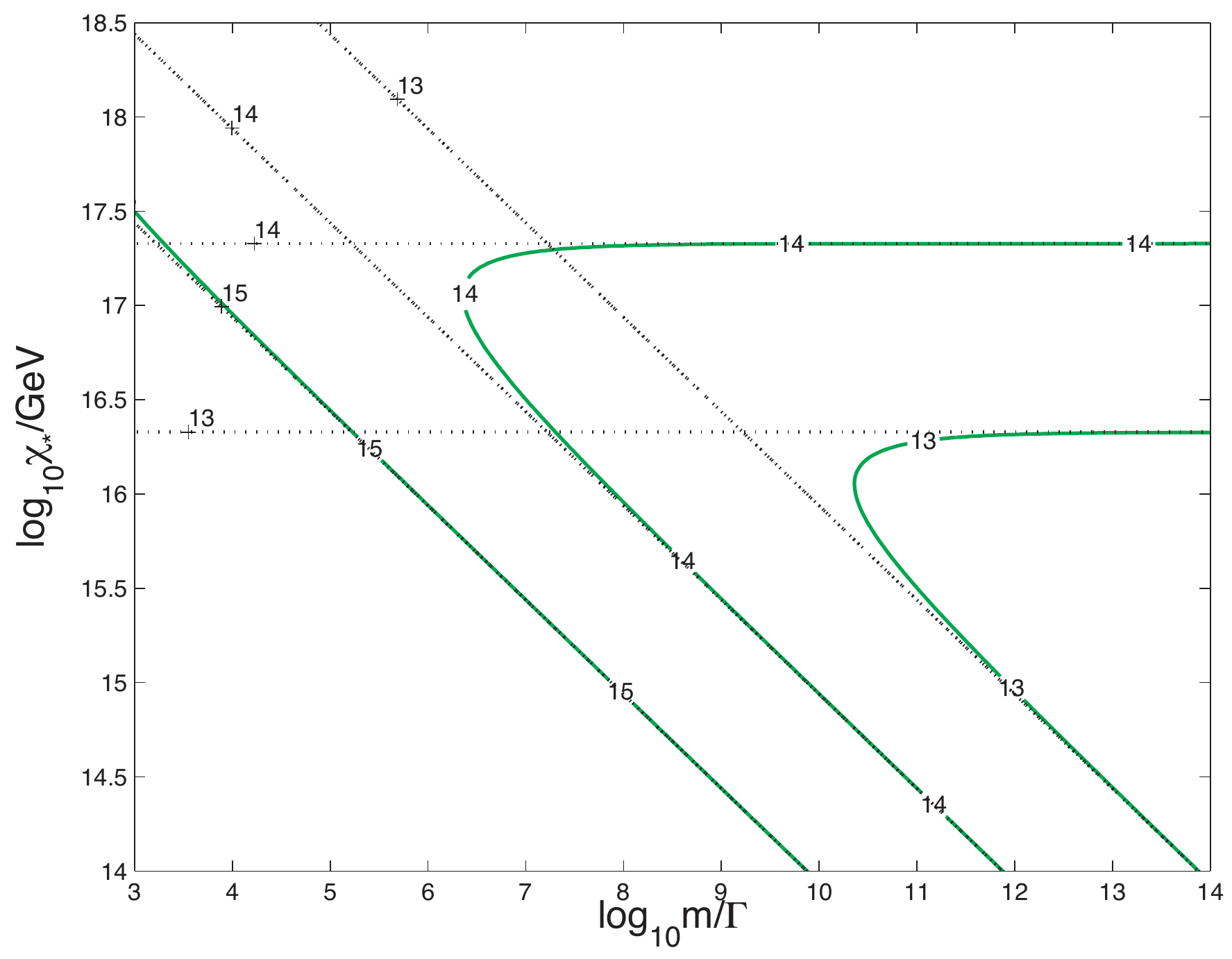}
\includegraphics[width=0.5\textwidth]{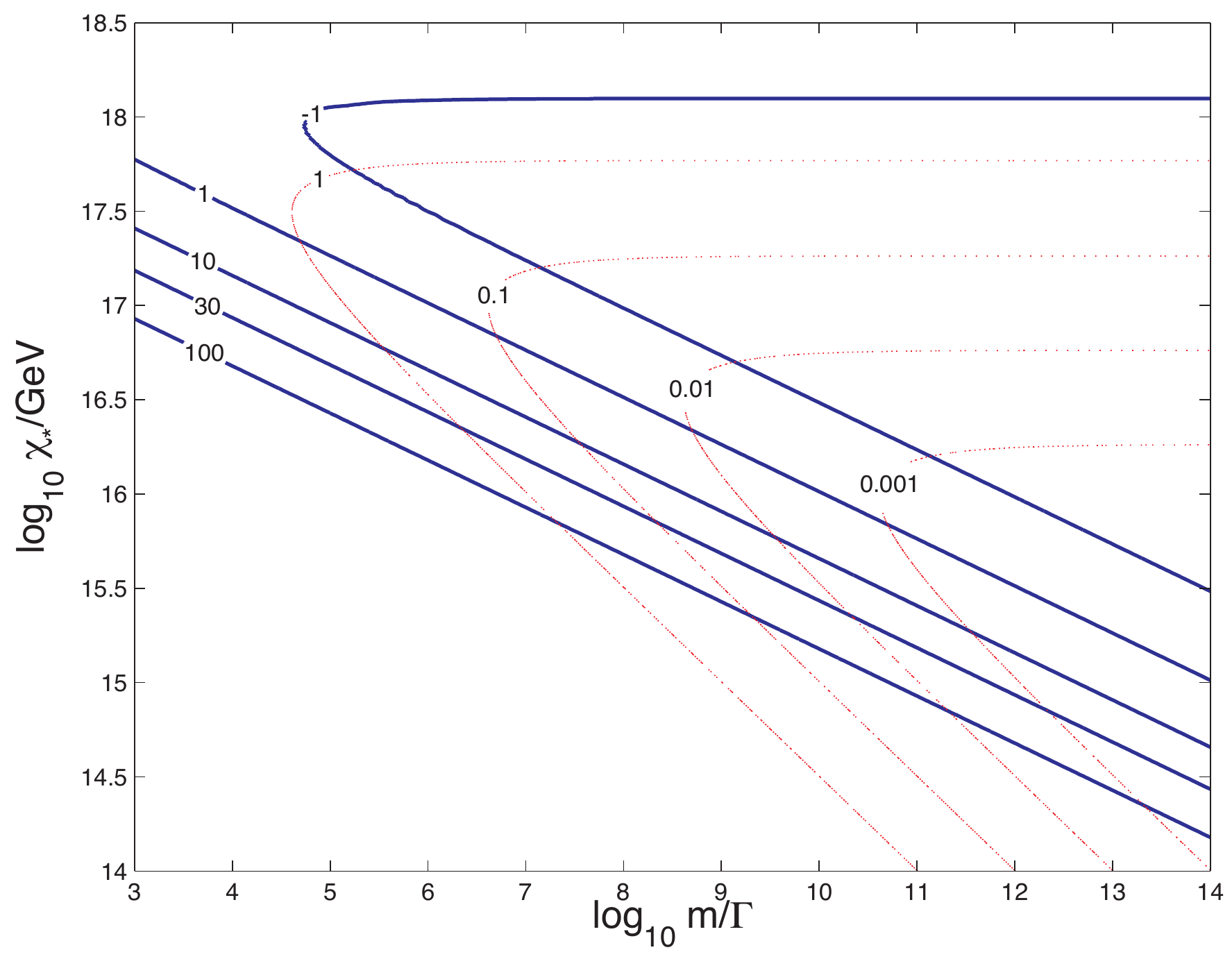}
\caption{Contour plots showing observational predictions for a curvaton field with quadratic potential as a function of the dimensionless decay rate, $\log_{10}(m/\Gamma)$, and the initial value of the curvaton, $\log_{10}(\chi_*/{\rm GeV})$. {\em Top:} Contour lines for the non-Gaussianity parameter $\fNL$ (in blue). The dotted black lines correspond to Eq. (\ref{fnlrsmall}). \ {\em Middle:} Contour lines for inflationary Hubble scale, $\log_{10}(H_*/{\rm GeV})$. The plotted contour lines correspond to $H_*=10^{13},10^{14},10^{15}$ GeV. The black dotted lines correspond to the 2 limits of Eq. (\ref{hlimits}). {\em Bottom:} Contour lines for both the non-Gaussianity parameter, $\fNL$, (blue thick solid line) and tensor-scalar ratio, $\rT$, (red dotted line).}
\label{fnlhcontourq}
\end{figure}

We show the results in Figure~\ref{compq} and~\ref{fnlhcontourq} for a quadratic curvaton potential. In this case we are able to compare our numerical result against an exact analytic expression while the curvaton density remains negligible during the radiation-dominated era. In this case the curvaton field is given by
\be
 \chi = \frac{\pi\chi_*}{2^{5/4}\Gamma(3/4)} \frac{J_{1/4}(mt)}{(mt)^{1/4}} \,.
 \ee
where $J_{1/4}(mt)$ is the Bessel function of the first kind of order $1/4$. This has the asymptotic solution $\chi\simeq1.023\chi_*\cos(mt-3\pi/8)/(mt)^{3/4}$, and substituting this into Eq.~(\ref{pdef}) gives
\be
 \label{pvchi*}
 p \simeq 1.046 \sqrt{\frac{m}{\Gamma}} \frac{\chi^2_*}{3\mPl^2} \,.
 \ee
We see from Figure~\ref{compq} that Eq.~(\ref{pvchi*}) gives an excellent approximation to the numerical results for $\chi_*\ll \mPl$.

Contour plots are given in Figure~\ref{fnlhcontourq} for the non-linearity parameter, $\fNL$, and the inflation Hubble scale, $H_*$, (and hence tensor-scalar ratio, $\rT$) for a non-self-interacting curvaton with a quadratic potential.

Given that the analytic result for $p(\chi_*)$ given in Eq.~(\ref{pvchi*}) is an excellent approximation, except for $\chi_*\sim\mPl$,
we deduce that $\chi_{\rm osc}=g(\chi_*)$ defined by Eq.~(\ref{posc}) is a linear function $g(\chi_*)\simeq \sqrt{2}\chi_*$. Thus the non-linearity parameter $\fNL$ is given in terms of $\R$ in Eq.~(\ref{fNLRlinear}).
We have two regimes for the transfer function $\R(p)$ given by Eq.~(\ref{Rp}). For $\chi_*\gg (\Gamma/m)^{1/4}\mPl$ we have $\pGupta\gg1$ and hence $\R\simeq1$, while for $\chi_*\ll (\Gamma/m)^{1/4}\mPl$ we have $\pGupta\ll1$ and hence $\R\simeq0.924\pGupta$.
Thus we find from Eq.~(\ref{fNLRlinear})
 \be
 \fNL \simeq
 \left\{
 \begin{array}{ll}
 -5/4 & {\rm for}\ \chi_* \gg (\Gamma/m)^{1/4}\mPl \\
 3.9 \sqrt{\frac{\Gamma}{m}} \frac{\mPl^2}{\chi_*^2} & {\rm for}\ \chi_* \ll (\Gamma/m)^{1/4}\mPl
 \end{array}
 \right.
 \,.
 \ee
Potentially observable levels of non-Gaussianity ($5<\fNL<100$) are found in a band of parameter space
 \be
 \chi_* \approx (1-4) \times 10^{17}\ {\rm GeV} \left( \frac{\Gamma}{10^{-6} m} \right)^{1/4} \,.
 \ee


The degeneracy between values of $\chi_*$ and $\Gamma/m$ which would be consistent with the same value of $\fNL$ is broken by a measurement of the scalar to tensor-ratio, $\rT$.
Substituting the approximation (\ref{pvchi*}) in Eq.~(\ref{hp}). We have
 \be
 H_* \simeq 4.7\times10^{-4} \frac{\chi_*}{\R(p)} \,,
 \ee
This yields two simple expressions for $H_*$ according to whether $\pGupta\gg1$ and hence $\R\simeq1$ or $\pGupta\ll1$ and hence $\R\simeq0.924\pGupta$. We thus have
 \be \label{hlimits}
 H_* \simeq
 \left\{
 \begin{array}{ll}
 4.7\times10^{-4} {\chi_*} & {\rm for}\ \chi_* \gg (\Gamma/m)^{1/4}\mPl \\
 1.5\times10^{-3} \sqrt{\frac{\Gamma}{m}} \frac{\mPl^2}{\chi_*} & {\rm for}\ \chi_* \ll (\Gamma/m)^{1/4}\mPl
 \end{array}
 \right.
 \,.
 \ee

Even a conservative bound on the tensor-scalar ratio such as $\rT<1$ thus places important bounds on the curvaton model parameters. Firstly there is the model-independent bound on the inflation Hubble scale, $H_*<2.7\times10^{14}$~GeV. In the case of a quadratic curvaton potential this imposes an upper bound on the value of the curvaton during inflation
 \be
 \chi_*<
 5.7 \times 10^{17}~{\rm GeV} \,,
 \ee
which is consistent with $\chi_*<\mPl$ required to use the analytic approximation (\ref{pvchi*}). We also find an upper bound on the dimensionless decay rate
 \be
 \frac{\Gamma}{m} < 0.023 \left( \frac{\chi_*}{\mPl} \right)^2 \,,
 \ee
and in any case $\Gamma<10^{-3}m$.
For example, for a TeV mass curvaton \cite{Enqvist:2010ky} we require $\Gamma<1$~GeV. More generally, if we require the curvaton to decay before primordial nucleosynthesis at a temperature of order $1$~MeV, we require $\Gamma>H_{\rm BBN}$ and hence $m>10^3H_{\rm BBN}$. On the other hand if the curvaton decays before decoupling of the lightest supersymmetric particle at a temperature of order $10$~GeV, we require $\Gamma>10^{-17}$~GeV and hence $m>10^{-14}$~GeV.

Bounds on the curvaton decay rate due to gravitational wave bounds were also studied recently in Ref.~\cite{Nakayama:2009ce}, who also considered the case where that curvaton oscillations begin immediately after inflation has ended at $H<m$.

We note that bounds on the tensor-scalar ratio rule out large regions of parameter space that would otherwise give rise to large non-Gaussianity.

A simultaneous measurement of primordial non-Gaussianity, $\fNL$, and primordial gravitational waves, $\rT$, for a non-self-interacting curvaton field with quadratic potential would determine both the energy scale of inflation, $H_*$, and the expectation value of the curvaton, $\chi_*$. It would also determine the dimensionless decay rate $\Gamma/m$, but not the absolute value of the mass and decay rate separately.
More optimistically, if the gravitational amplitude was large enough to determine the tensor tilt, $n_T$ and hence $\epsilon$, the scale dependence of the scalar spectrum would determine the curvaton mass:
\be
m_\chi^2 = 3\eta_\chi H_*^2 \simeq \frac32 (n_\zeta-1-n_T) \frac{r_T}{2.0\times 10^7} \mPl^2
\ee

However once $\epsilon$ is known then from Eq.~(\ref{zetastar}) we also know the curvature perturbation due to inflaton perturbations during slow-roll inflation: $\P_{\zeta*}=(\rT/16\epsilon)\P_\zeta$. If $\epsilon\leq 0.02$, as is commonly assumed, then our assumption that the inflaton perturbations are negligible is no longer valid for $\rT\sim0.3$. In this case we need to consider a mixed inflaton-curvaton model. This inflaton-curvaton model has a much richer phenomenology \cite{Langlois:2004nn,Moroi:2005kz,Moroi:2005np,Ichikawa:2008iq,Suyama:2010uj} and we leave the study of the combined non-Gaussianity and gravitational wave bounds in this scenario to future work. Otherwise we must assume $\epsilon$ is sufficiently large that the inflaton-generated perturbations remain negligible.

\subsection{Self-interacting curvaton}

We have seen that non-linear field evolution due to gravitational back-reaction of the curvaton field with a quadratic potential is limited to large initial values $\chi_*\sim\mPl$ which are incompatible with bounds on the tensor-scalar ratio in the curvaton scenario with a quadratic potential. However significant non-linear field evolution may arise from self-interactions of the curvaton field, due to deviations from a purely quadratic potential. Polynomial self-interaction terms of the form $V_{\rm int}\propto \chi^n$ where $n\geq4$ have been shown to have a large effect on observational predictions in some regions of parameter space \cite{Huang:2008zj,Enqvist:2009zf,Enqvist:2009ww}.

Rather than choose a monomial correction term, we choose a functional form that leads to significant corrections at a specified mass scale. In particular we are motivated by axion type potentials where the curvaton field has a natural range, $f$. Thus we consider a cosine-type potential, with a smaller mass effective mass for $\chi_*\sim f$ and a hyperbolic-cosine potential which has a much larger mass for $\chi_*\sim f$. In both cases the corrections lead to a finite range $\chi_*\sim f$ for the initial curvaton field.

\subsubsection{Cosine potential}

We consider an axion-type potential for a weakly-broken $U(1)$-symmetry\cite{Lyth:2001nq,Dimopoulos:2003az}
\be
 \label{Vcos}
V(\chi) = M^4 \left( 1- \cos \left( \frac{\chi}{f} \right) \right) \simeq \frac12 m^2\chi^2 - \frac{1}{24} \frac{m^2\chi^4}{f^2} + \ldots \,,
\ee
where $m^2=M^4/f^2\ll M^2$ and we have an additional model parameter corresponding to the mass scale $f\gg M$ which determines the relative importance of self-interaction terms at a given curvaton field value. It also determines a natural expectation value for the curvaton field, $\chi_*\sim f$. In the following we assume $f<\mPl$.

In Figure~\ref{compq} we show the numerical solution for $\pFW$ as a function of $\chi_*$, corresponding to $\pGupta$ for a fixed value of $m/\Gamma$. As expected we see that for $\chi_*\ll f$ we recover the analytic result (\ref{pvchi*}) as the potential is effectively quadratic and self-interactions have a negligible effect. For larger values of $\chi_*$, the potential becomes flatter than the corresponding quadratic potential and we see that $\pFW$, and hence $\pGupta$, can become much larger than would be obtained for a quadratic correction. Note that the potential (\ref{Vcos}) is periodic and we can identify $\pFW(\chi_*+\pi f/2)=\pFW(\pi f/2-\chi_*)$.

\begin{figure}
\centering
\includegraphics[width=0.5\textwidth]{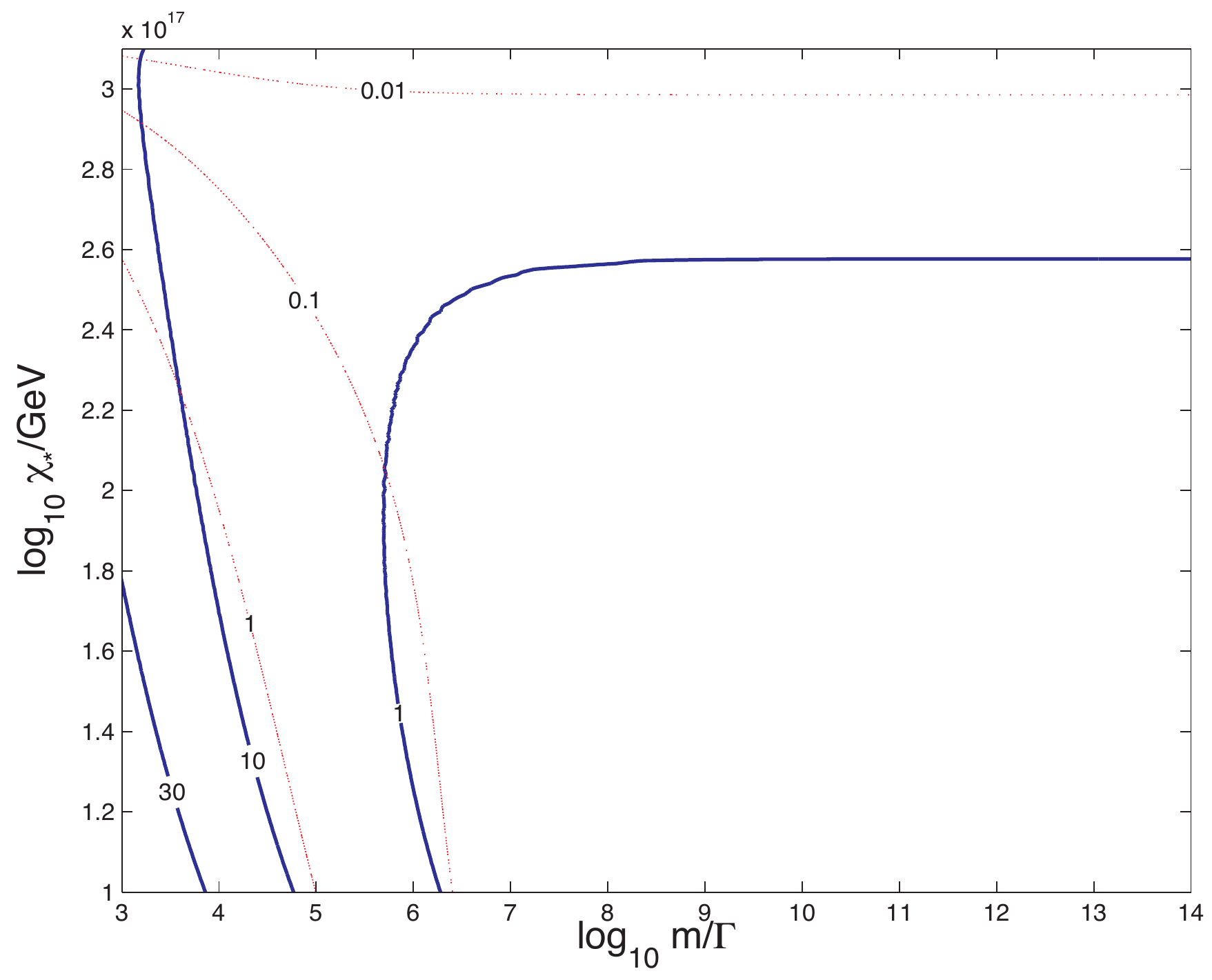}
\includegraphics[width=0.5\textwidth]{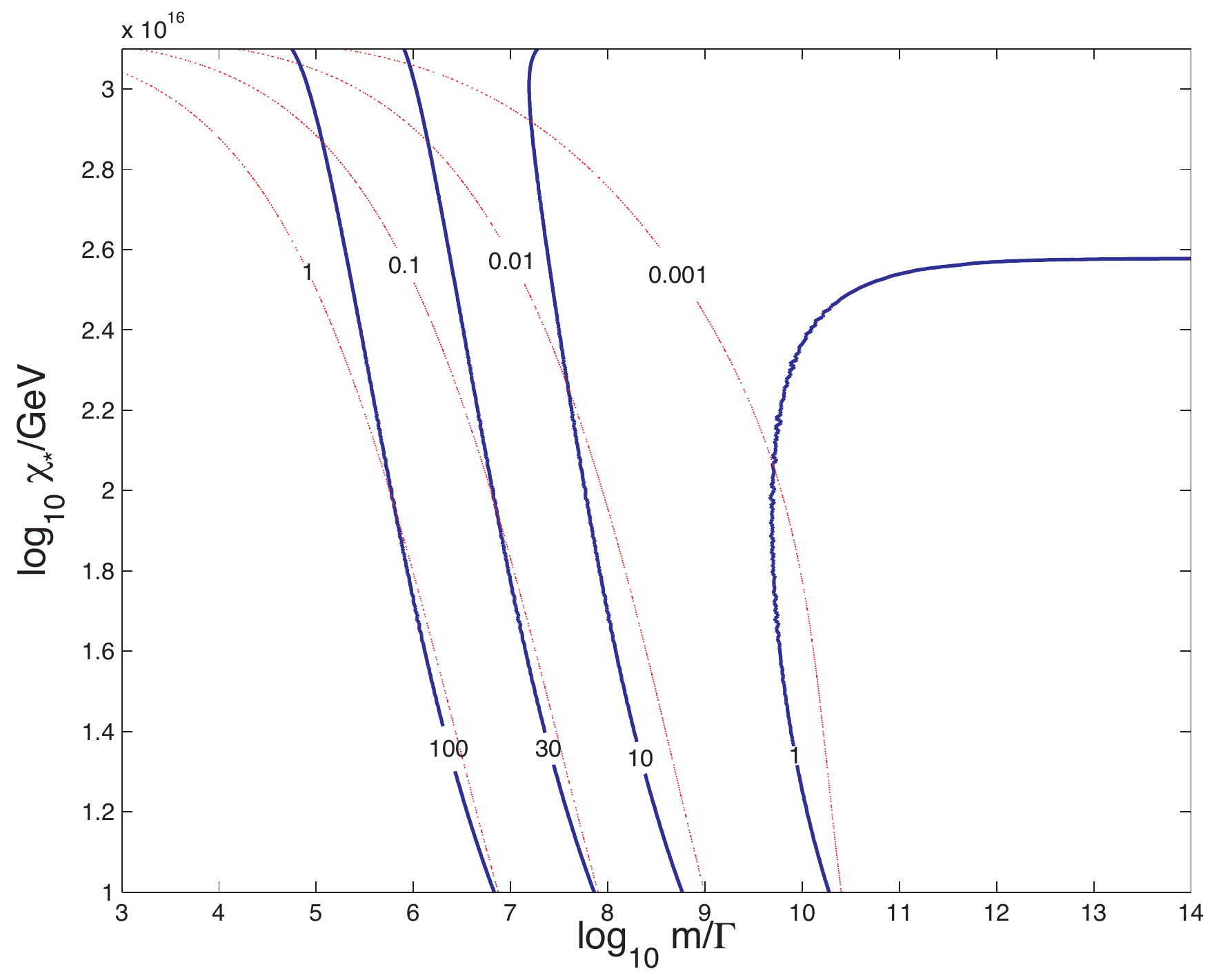}
\includegraphics[width=0.5\textwidth]{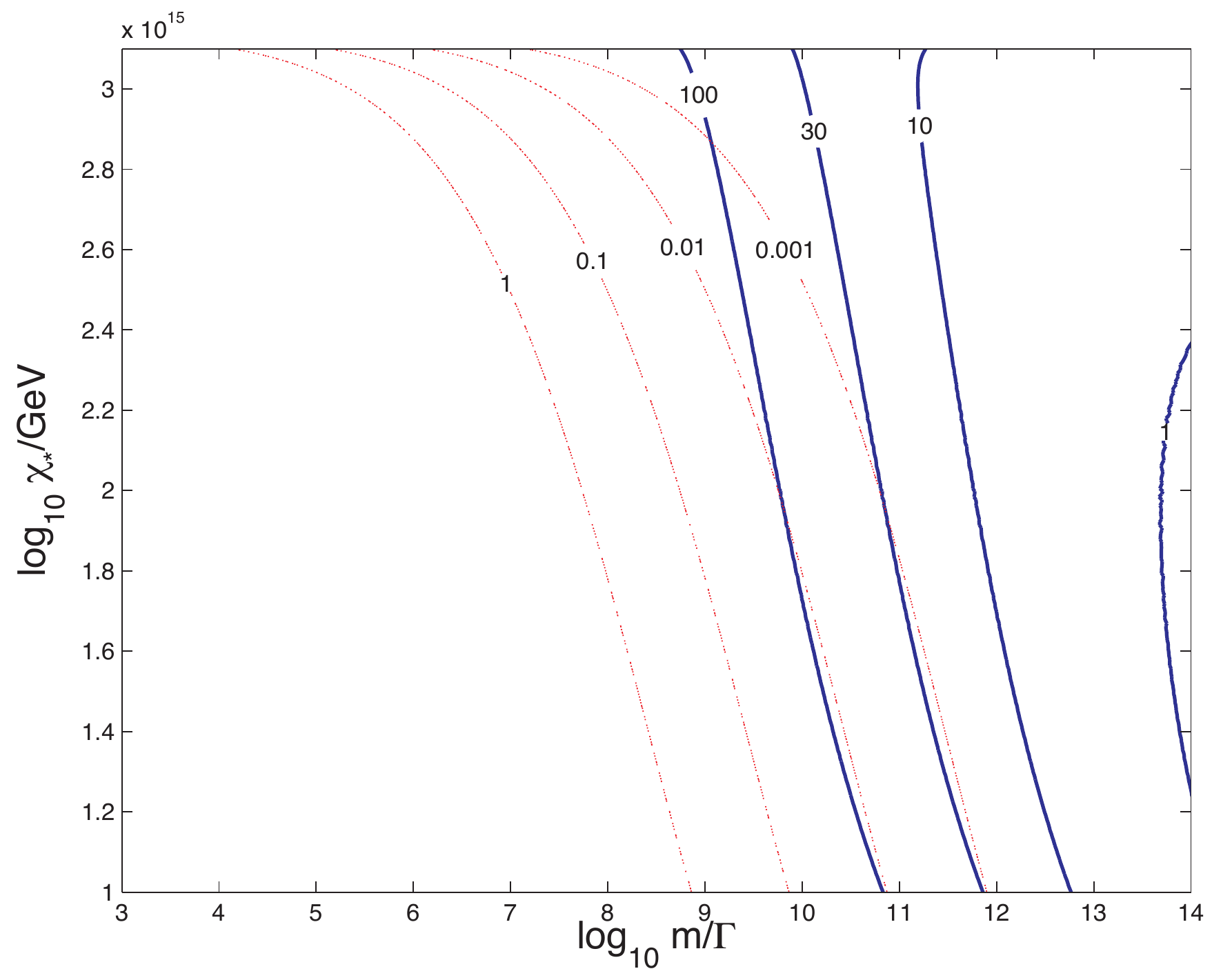}
\caption{Contour plots showing observational predictions for a curvaton field with a cosine type potential. The three plots show, from top to bottom, observational parameters for cosine potentials with $f=10^{17}$GeV, $f=10^{16}$GeV and $f=10^{15}$GeV, respectively,
as a function of the dimensionless decay rate, $\log_{10}(m/\Gamma)$, and the initial value of the curvaton, $\chi_*/{\rm GeV}$.
Thick solid blue contour lines show bispectrum amplitude, $\fNL$, decreasing from left to right. Dotted red contour lines show the tensor-scalar ratio $\rT$, also decreasing from left to right.
}
\label{fnlhcontourcos}
\end{figure}

We show numerical predictions for the non-Gaussianity parameter, $\fNL$, and the tensor-scalar ratio, in Figure~\ref{fnlhcontourcos}. Non-linear evolution of the field becomes important for $\chi_*\sim f$. In particular we see that an upper bound on the tensor-scalar ratio no longer places an upper bound on the decay rate $\Gamma/m$ as we approach the top of the potential, i.e., as $\chi_*\to \pi f$.

Modest, positive values of the non-linearity parameter, $1<\fNL<10$, become possible even if the curvaton dominates the energy density when it decays ($p>1$) if $\chi_*>2.5 f$, but we never find very large values of $\fNL>100$. Because $g''>0$ in Eq.~(\ref{fNLR}) we have $\fNL>-5/4$, as in the case of a quadratic potential, and we never find large negative values of $\fNL$ for a cosine-type potential.

\subsubsection{Hyperbolic-cosine potential}

Non-linearity of the cosine potential (\ref{Vcos}) yields a flat potential with small effective mass during inflation for $\chi_*\sim f$. To consider the effect of self-interactions leading to a larger effective mass we consider a hyperbolic cosine potential which becomes an exponential function of the curvaton field at large field values, as may be expected due to supergravity corrections.
\be
V(\chi) = M^4 \left( \cosh \left( \frac{\chi}{f} \right) - 1 \right) \simeq \frac12 m^2\chi^2 + \frac{1}{24} \frac{m^2\chi^4}{f^2} + \ldots \,.
\ee
In the following we assume $f<\mPl$. As in the case of the cosine potential, this also yields a natural range for $\chi_*\sim f$. In the case of a hyperbolic potential, the field becomes heavy relative to the Hubble scale and evolves rapidly for values of $\chi_*$ much larger than $f$. In particular the requirement that the curvaton have an effective mass less than $0.1H$ at the start of our numerical solutions imposes the constraint $\chi_*<5f$.

In Figure~\ref{compq} we show the numerical solution for $\pGupta$ as a function of $\chi_*$ for a fixed value of $m/\Gamma$. As expected we see that for $\chi_*\ll f$ we recover the analytic result (\ref{pvchi*}) when the potential is effectively quadratic. However for the a hyperbolic potential we see that due to the steeper potential the effective energy density when the curvaton decays, determined by the parameter $\pGupta$, becomes less than the quadratic case for $\chi_*\sim f$.

\begin{figure}
\centering
\includegraphics[width=0.5\textwidth]{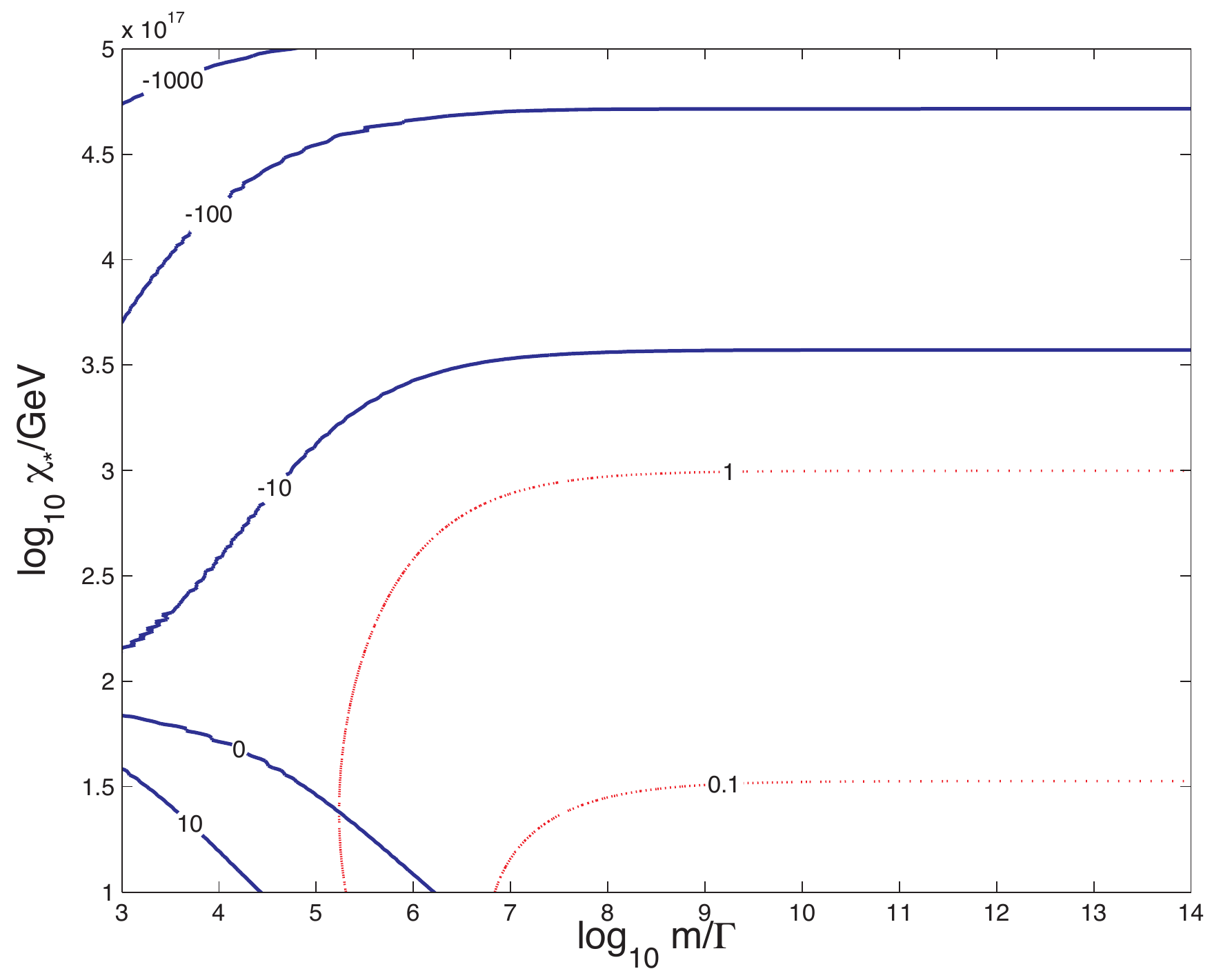}
\includegraphics[width=0.5\textwidth]{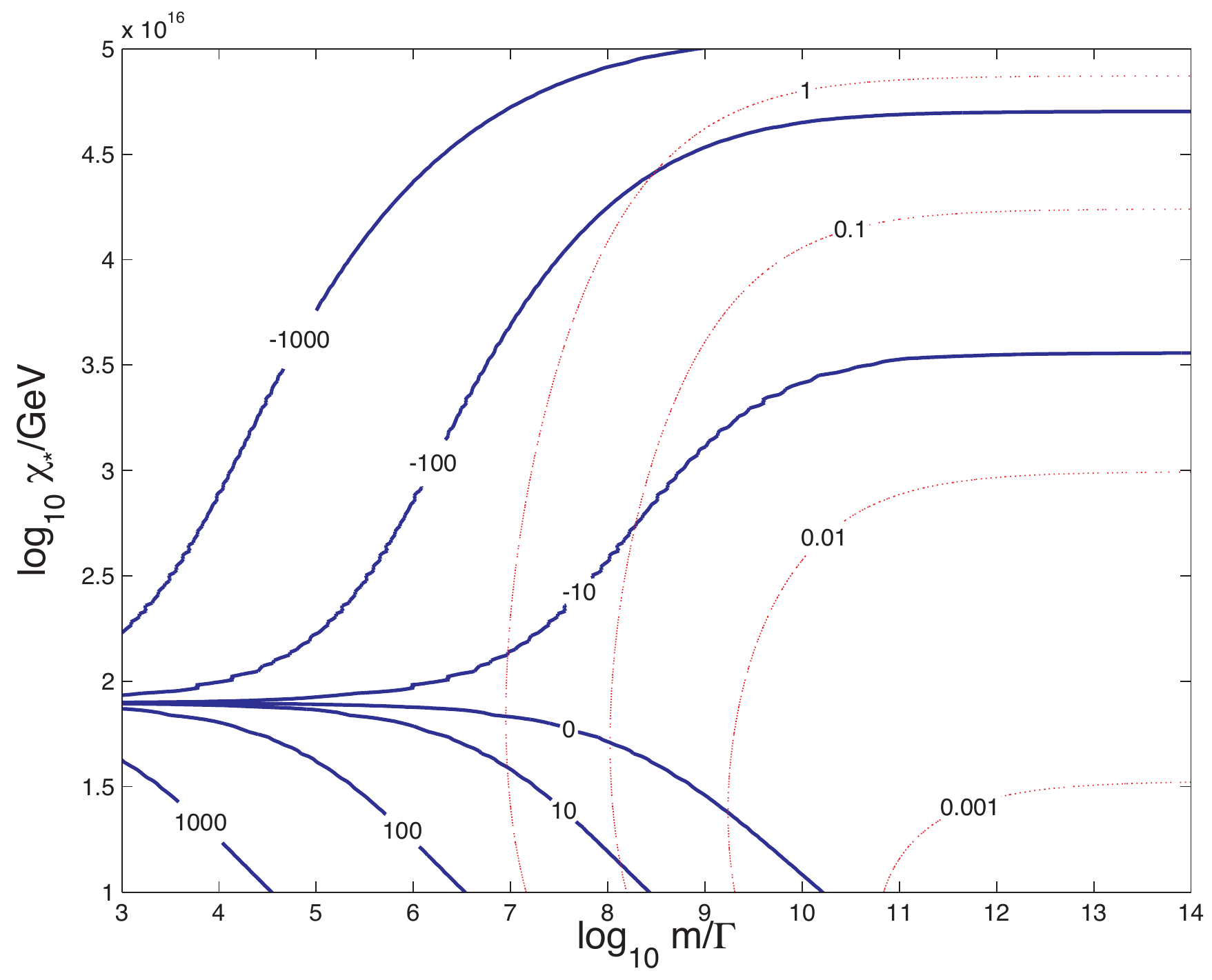}
\includegraphics[width=0.5\textwidth]{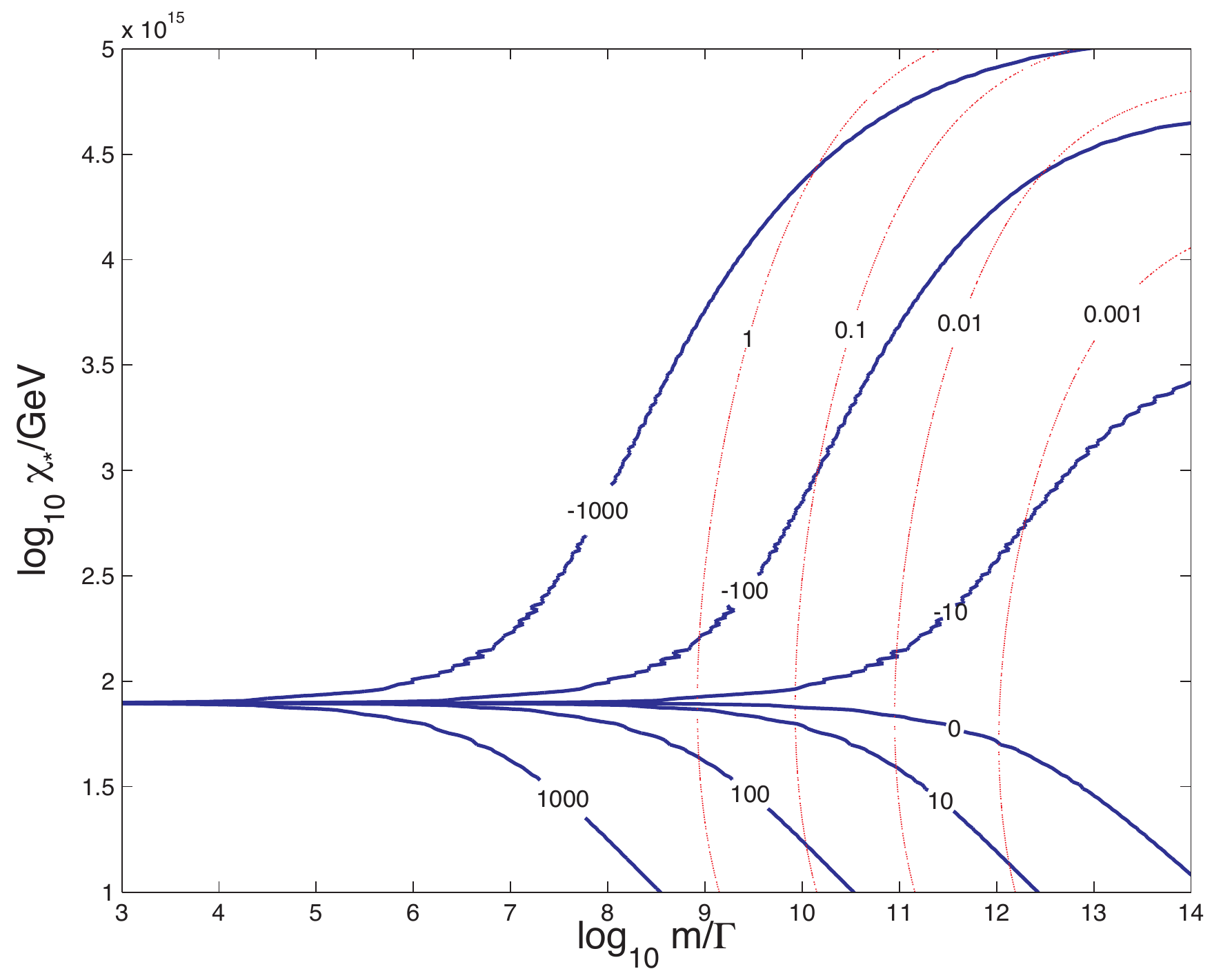}
\caption{Contour plots showing observational predictions for a curvaton field with a hyperbolic-cosine-type potential. The three plots show, from top to bottom, observational parameters for potentials with $f=10^{17}$GeV, $f=10^{16}$GeV and $f=10^{15}$GeV, respectively,
as a function of the dimensionless decay rate, $\log_{10}(m/\Gamma)$, and the initial value of the curvaton, $\chi_*/{\rm GeV}$.
Thick solid blue contour lines show bispectrum amplitude, $\fNL$, increasing from top to bottom.
Dotted red contour lines show the tensor-scalar ratio, $\rT$, decreasing from left to right.
}
\label{fnlhcontourcosh}
\end{figure}

We show numerical predictions for the non-Gaussianity, $\fNL$, and the inflation Hubble scale, $H_*$, (and hence the tensor-scalar ratio) in Figure~\ref{fnlhcontourcosh}. The non-linear correction $g''g/g^{\prime2}$ in Eq.~(\ref{fNLR}) becomes negative for $\chi_*\sim f$ and we can obtain large negative values of $\fNL$.

However we find that the bound on the tensor-scalar plays an important role. Regions of parameter space which yield large negative $\fNL$ also give large tensor-scalar ratios. In regions where $\pGupta\gg1$ and the curvaton dominates when it decays we have $\R\sim1$ and both the tensor-scalar ratio and the non-linearity parameters become functions solely of $\chi_*$.
In this regime, we have, from Eqs.~(\ref{rTp}) and~(\ref{fNLR})
\begin{eqnarray}
\rT
 \simeq \frac{9}{2} \left( \frac{g}{g'\mPl} \right)^2 \, ,\\
 \fNL \simeq \frac{5}{4} \left( \frac{g''g}{g^{\prime2}} \right) \,,
\end{eqnarray}
which are both clearly functions of $\chi_*$.
Indeed formally we can eliminate $g(\chi_*)$ and its derivatives in order to write
\begin{equation}
 \fNL \simeq - \sqrt{\frac{25}{72}} \mPl \left( \sqrt{\rT} \right)^\prime \,.
\end{equation}
Hence the contours of equal values of both $\rT$ and $\fNL$ become horizontal on the right-hand-side of Figure~\ref{fnlhcontourcosh}. For example, with $f=10^{16}$~GeV a weak bound on the tensor-scalar ratio of $\rT<1$ requires $\fNL>-1000$. A stronger bound $\rT<0.1$ requires $\fNL>-100$.

Of course $(\sqrt{\rT})^\prime$ is not an observable parameter, but if we assume that $\sqrt{\rT}$ is a smooth function of $\chi_*/f$ we can estimate $(\sqrt{\rT})^\prime\sim (\sqrt{\rT})/f$ and hence
\begin{equation}
 \fNL \sim - \frac{\mPl}{f} \sqrt{\rT} \,.
\end{equation}
This semi-empirical relation appears to hold for sufficiently small $\Gamma/m$ and it would be interesting to see if this is also the case for the polynomial correction terms \cite{Enqvist:2009zf,Enqvist:2009ww}.


Unlike the case of a cosine-type potential we still have a strict upper bound on the decay rate, as in the case of a purely quadratic potential. Thus, although there are regions of parameter space for $\chi_*\sim f$, where the non-Gaussianity can be small even if the curvaton is subdominant when it decays, $\pGupta\ll1$, we find that these regions correspond to large values for the tensor-scalar ratio and are excluded by bounds on primordial gravitational waves.




\section{Discussion and Conclusions}


In this work we have investigated the numerical evolution of a curvaton field from its overdamped regime after inflation until it decays into radiation. We have shown how measurement of both the non-linearity parameter, $\fNL$, and the tensor-to-scalar ratio, $\rT$, provide complementary constraints on the model parameters. We did this for three different curvaton potentials: the quadratic potential, axion-like cosine potentials and hyperbolic potentials. As expected both the cosine and the hyperbolic potentials recover the quadratic regime when $\chi_*\ll f$.

For the simplest quadratic potential for the curvaton, bounds on the tensor-to-scalar ratio place an upper bound on the dimensionless decay rate, ruling out large regions of parameter space that would yield a large primordial non-Gaussianity in the distribution of scalar perturbations.
Simultaneous measurement of both the non-linearity parameter, $\fNL$, and the tensor-to-scalar ratio, $\rT$, can determine both the expectation value of the field during inflation, $\chi_*$, and the dimensionless decay rate, $\Gamma/m$.

In the conventional inflaton scenario for the origin of structure we have three free parameters: the inflation scale $H_*$ and two slow-roll parameters, $\epsilon$ and $\eta_{\phi}$. These can be determined by power of the primordial scalar perturbations, $\P_{\zeta}$, the tensor perturbations, $\P_{T}$, and the spectral index of the scalar spectrum, $n_{\zeta}$. The spectral index of the tensor spectrum, if measurable, would give a valuable consistency check \cite{Baumann:2008aq}. Another important consistency condition for canonical, slow-roll inflation is that the primordial density perturbations should be Gaussian and the non-linearity parameter, $\fNL$, should be much less that unity \cite{Maldacena:2002vr}.

In the curvaton scenario with a simple quadratic potential we have 5 free parameters: the inflation scale $H_*$, the expectation value of the curvaton during inflation $\chi_*$, the decay rate of the curvaton relative to its mass, $\Gamma/m$, and the slow roll parameters $\epsilon$ and $\eta_{\chi}=m_*^2/3H_*^2$.
For a curvaton, we find that $H_*$, $\chi_*$ and $\Gamma/m$ are determined by the primordial scalar perturbations, $\P_{\zeta}$, the tensor perturbations, $\P_{T}$, and the non-linearity parameter, $\fNL$, but the mass and decay rate of the curvaton are not separately determined. The two slow-roll parameters $\epsilon$ and $\eta_{\chi}$ are then determined by the two spectral indices $n_{\zeta}$, and $n_{T}$.

Another natural observable in the curvaton model is the scale dependence of the non-linearity parameter, defined as \cite{Byrnes:2009pe}
\be
n_{\fNL} \equiv \frac{d \ln|\fNL|}{d\ln k}
\ee
In the curvaton scenario this is given by a simple expression \cite{Byrnes:2010xd,Huang:2010cy}
\be
 n_{\fNL} = \eta_3 \frac{g}{\mPl g'} \frac{5}{4\R\fNL} \,.
 \ee
where we define $\eta_3 \equiv \mPl^3 V'''/V$. This can be rewritten in terms of observable quantities and $\eta_3$
 \be
 n_{\fNL} = \eta_3 \frac{5}{12\sqrt{2}} \frac{\sqrt{r_T}}{\fNL} \,.
 \ee
Thus it offers the possibility of testing the curvaton self interactions. Future observations may be able to detect $|\fNL n_{\fNL}|> 5$ \cite{Sefusatti:2009xu}, corresponding to $|\eta_3|\sqrt{\rT}>17$. For the quadratic potential we have the consistency condition $n_{\fNL}=0$.

Deviations from a quadratic potential introduce at least one further model parameter, $f$, corresponding to the mass scale associated with the non-linear corrections. This leads to a degeneracy in model parameters consistent with the five observables $\P_{\zeta}$, $\P_{T}$, $\fNL$, $n_{\zeta}$ and $n_{T}$, but this can be broken by a measurement of $n_{\fNL}$.

In the case of a cosine-type curvaton potential the self interaction corrections became important near the top of the potential, i.e., when $\chi_* \sim \pi f$ \cite{Huang:2010cy} and the tensor-to-scalar ratio no longer places an upper bound on $\Gamma/m$. As for a quadratic curvaton, we still find $\fNL>-5/4$ and hence any large non-Gaussianity, $|\fNL|\gg1$, has positive $\fNL$. But for $\chi_*\sim f$ we have $\eta_3 \sim -(\mPl/f^3)<0$, and if $f$ is well below the Planck scale there could be strong scale dependence.


In the case of a hyperbolic-type potential $\fNL$ can become large and negative, for $\chi_*\sim f$. However the tensor-to-scalar ratio again plays an important role, in this case placing a lower bound on $\fNL$, e.g., $\fNL>-100$ for $\rT<0.1$ when $f=10^{16}$~GeV. In this regime we find $\eta_3 \sim (\mPl/f^3)>0$, which can be large, leading to strong scale dependence for $f\ll \mPl$, with $n_{\fNL}<0$ for $\fNL<0$.

Running of either the scalar tilt, $\alpha_\zeta$, or the non-linearity, $\alpha_{\fNL}$ \cite{Huang:2010cy}, yields additional information about the higher derivatives of the potential, and in particular curvaton-inflaton interactions which we have assumed are negligible in our analysis.

Significant non-Gaussianity in the primordial perturbations opens up the possibility to extract information from the higher-order correlations in the scalar spectrum, such as the trispectrum \cite{Byrnes:2006vq,Sasaki:2006kq,Enqvist:2008gk,Huang:2008bg,Enqvist:2009ww}
\bea
T_{\zeta}(k_1,k_2,k_3,k_4) = \frac{54}{25} \gNL \left[ P_{\zeta}(k_2)P_{\zeta}(k_3)P_{\zeta}(k_4) +3 ~ {\rm perms} \right] + \frac{36}{25} \fNL^2 \left[P_{\zeta}(k_{13})P_{\zeta}(k_3) P_{\zeta}(k_4) + 11~{\rm perms} \right] \,.
\eea
which are sensitive to higher-order derivatives of the expansion history with respect to the curvaton field value during inflation through $\gNL= (25/54)N'''/N'^3$.
Differentiating Eq.~(\ref{fNLRused}) we obtain
\be
 \label{gNLRused}
\gNL= \frac {25}{24} \left[ \frac{\R''}{\R^3}\frac{g^2}{g'^2}+2\frac{\R'}{\R^3}\left(\frac{g^2g''}{g'^3}-\frac{g}{g'}\right)+\frac 1 {\R^2} \left(\frac{g^2g'''}{g'^3}-3\frac{gg''}{g'^2}+2\right)\right] \,.
\ee
which using the sudden-decay approximation can be written as \cite{Sasaki:2006kq,Byrnes:2006vq}
 \be
 \gNL = \frac{25}{54} \left[ \frac{9}{4\R^2} \left( \frac{g^2g'''}{g^{\prime3}} + 3\frac{gg''}{g^{\prime2}} \right) - \frac9\R \left( 1 + \frac{gg''}{g^{\prime2}} \right) + \frac12 \left( 1 -9 \frac{gg''}{g^{\prime2}} \right) +10\R +3\R^2 \right]
 \,.
 \ee
$\gNL$ and its scale dependence $n_{\gNL}$ \cite{Byrnes:2010ft,Byrnes:2010xd}, thus provide additional observable parameters which then offer consistency conditions for generalised curvaton models such as the cosine or hyperbolic potentials.
In practice we require more accurate numerical simulations than those used in this work to reliably determine the required higher-derivatives with respect to the initial field value across the range of model parameters used in this paper and we leave this for future work.

\acknowledgements{The authors are grateful to Hooshyar Assadullahi and Tomo Takahashi for helpful comments. JF is supported by ÒFunda\c c\~ao para a Ci\^encia e a Tecnologia (Portugal), fellowships reference number SFRH/BD/40150/2007, and by the GCOE program at Kyoto University, ÒThe Next Generation of Physics, Spun from Universality and EmergenceÓ. DW is supported by STFC grant ST/H002774/1. The authors are grateful to the Yukawa Institute for Theoretical Physics for their hospitality and the organisers of the YITP workshop, YITP-T-09-05, ``The Non-Gaussian Universe''.}

{}
\end{document}